\documentclass[10pt,a4paper,onecolumn]{article}
\usepackage{marginnote}
\usepackage{graphicx}
\usepackage{xcolor}
\usepackage{authblk,etoolbox}
\usepackage{titlesec}
\usepackage{calc}
\usepackage{tikz}
\usepackage{hyperref}
\hypersetup{colorlinks,breaklinks=true,
            urlcolor=[rgb]{0.0, 0.5, 1.0},
            linkcolor=[rgb]{0.0, 0.5, 1.0}}
\usepackage{caption}
\usepackage{tcolorbox}
\usepackage{amssymb,amsmath}
\usepackage{ifxetex,ifluatex}
\usepackage{seqsplit}
\usepackage{xstring}

\usepackage{float}
\let\origfigure\figure
\let\endorigfigure\endfigure
\renewenvironment{figure}[1][2] {
    \expandafter\origfigure\expandafter[H]
} {
    \endorigfigure
}

\usepackage{fixltx2e} 
\usepackage[
  backend=biber,
]{biblatex}
\bibliography{paper.bib}

\let\textttOrig=\texttt
\def\texttt#1{\expandafter\textttOrig{\seqsplit{#1}}}
\renewcommand{\seqinsert}{\ifmmode
  \allowbreak
  \else\penalty6000\hspace{0pt plus 0.02em}\fi}

\makeatletter
\let\href@Orig=\href
\def\href@Urllike#1#2{\href@Orig{#1}{\begingroup
    \def\Url@String{#2}\Url@FormatString
    \endgroup}}
\def\href@Notdoi#1#2{\def\tempa{#1}\def\tempb{#2}%
  \ifx\tempa\tempb\relax\href@Urllike{#1}{#2}\else
  \href@Orig{#1}{#2}\fi}
\def\href#1#2{%
  \IfBeginWith{#1}{https://doi.org}%
  {\href@Urllike{#1}{#2}}{\href@Notdoi{#1}{#2}}}
\makeatother

\newlength{\cslhangindent}
\newlength{\csllabelwidth}
\newenvironment{CSLReferences}[3] 
 {
  \setlength{\parindent}{0pt}
  \ifodd #1 \everypar{\setlength{\hangindent}{\cslhangindent}}\ignorespaces\fi
  \ifnum #2 > 0
  \setlength{\parskip}{#2\baselineskip}
  \fi
 }%
 {}
\usepackage{calc}

\usepackage[top=3.5cm, bottom=3cm, right=1.5cm, left=1.0cm,
            headheight=2.2cm, reversemp, includemp, marginparwidth=4.5cm]{geometry}

\titleformat{\section}
  {\normalfont\sffamily\Large\bfseries}
  {}{0pt}{}
\titleformat{\subsection}
  {\normalfont\sffamily\large\bfseries}
  {}{0pt}{}
\titleformat{\subsubsection}
  {\normalfont\sffamily\bfseries}
  {}{0pt}{}
\titleformat*{\paragraph}
  {\sffamily\normalsize}

\usepackage{fancyhdr}
\makeatletter
\let\ps@plain\ps@fancy
\fancyheadoffset[L]{4.5cm}
\fancyfootoffset[L]{4.5cm}

\definecolor{linky}{rgb}{0.0, 0.5, 1.0}

\newtcolorbox{repobox}
   {colback=red, colframe=red!75!black,
     boxrule=0.5pt, arc=2pt, left=6pt, right=6pt, top=3pt, bottom=3pt}

\newcommand{\ExternalLink}{%
   \tikz[x=1.2ex, y=1.2ex, baseline=-0.05ex]{%
       \begin{scope}[x=1ex, y=1ex]
           \clip (-0.1,-0.1)
               --++ (-0, 1.2)
               --++ (0.6, 0)
               --++ (0, -0.6)
               --++ (0.6, 0)
               --++ (0, -1);
           \path[draw,
               line width = 0.5,
               rounded corners=0.5]
               (0,0) rectangle (1,1);
       \end{scope}
       \path[draw, line width = 0.5] (0.5, 0.5)
           -- (1, 1);
       \path[draw, line width = 0.5] (0.6, 1)
           -- (1, 1) -- (1, 0.6);
       }
   }

\patchcmd{\@maketitle}{center}{flushleft}{}{}
\patchcmd{\@maketitle}{center}{flushleft}{}{}
\patchcmd{\@maketitle}{\LARGE}{\LARGE\sffamily}{}{}
\def\maketitle{{%
  
  \AB@maketitle}}
\makeatletter
\renewcommand\AB@affilsepx{ \protect\Affilfont}
\renewcommand\AB@affilnote[1]{{\bfseries #1}\hspace{3pt}}
\renewcommand{\affil}[2][]%
   {\newaffiltrue\let\AB@blk@and\AB@pand
      \if\relax#1\relax\def\AB@note{\AB@thenote}\else\def\AB@note{#1}%
        \setcounter{Maxaffil}{0}\fi
        \begingroup
        \let\href=\href@Orig
        \let\texttt=\textttOrig
        \let\protect\@unexpandable@protect
        \def\thanks{\protect\thanks}\def\footnote{\protect\footnote}%
        \@temptokena=\expandafter{\AB@authors}%
        {\def\\{\protect\\\protect\Affilfont}\xdef\AB@temp{#2}}%
         \xdef\AB@authors{\the\@temptokena\AB@las\AB@au@str
         \protect\\[\affilsep]\protect\Affilfont\AB@temp}%
         \gdef\AB@las{}\gdef\AB@au@str{}%
        {\def\\{, \ignorespaces}\xdef\AB@temp{#2}}%
        \@temptokena=\expandafter{\AB@affillist}%
        \xdef\AB@affillist{\the\@temptokena \AB@affilsep
          \AB@affilnote{\AB@note}\protect\Affilfont\AB@temp}%
      \endgroup
       \let\AB@affilsep\AB@affilsepx
}
\makeatother

\renewcommand\Affilfont{\sffamily\small\mdseries}
\ifnum 0\ifxetex 1\fi\ifluatex 1\fi=0 
  \usepackage[T1]{fontenc}
  \usepackage[utf8]{inputenc}

\else 
  \ifxetex
    \usepackage{mathspec}
    \usepackage{fontspec}

  \else
    \usepackage{fontspec}
  \fi
  \defaultfontfeatures{Ligatures=TeX,Scale=MatchLowercase}

\fi
\IfFileExists{upquote.sty}{\usepackage{upquote}}{}
\IfFileExists{microtype.sty}{%
\usepackage{microtype}
\UseMicrotypeSet[protrusion]{basicmath} 
}{}

\usepackage{hyperref}
\hypersetup{unicode=true,
            pdftitle={exoplanet: Gradient-based probabilistic inference for exoplanet data \& other astronomical time series},
            pdfborder={0 0 0},
            breaklinks=true}
\let\addcontentslineOrig=\addcontentsline
\def\addcontentsline#1#2#3{\bgroup
  \let\texttt=\textttOrig\addcontentslineOrig{#1}{#2}{#3}\egroup}
\let\markbothOrig\markboth
\def\markboth#1#2{\bgroup
  \let\texttt=\textttOrig\markbothOrig{#1}{#2}\egroup}
\let\markrightOrig\markright
\def\markright#1{\bgroup
  \let\texttt=\textttOrig\markrightOrig{#1}\egroup}

\usepackage{graphicx,grffile}
\makeatletter
\def\maxwidth{\ifdim\Gin@nat@width>\linewidth\linewidth\else\Gin@nat@width\fi}
\def\maxheight{\ifdim\Gin@nat@height>\textheight\textheight\else\Gin@nat@height\fi}
\makeatother
\setkeys{Gin}{width=\maxwidth,height=\maxheight,keepaspectratio}
\IfFileExists{parskip.sty}{%
\usepackage{parskip}
}{
\setlength{\parindent}{0pt}
\setlength{\parskip}{6pt plus 2pt minus 1pt}
}
\providecommand{\tightlist}{%
  \setlength{\itemsep}{0pt}\setlength{\parskip}{0pt}}
\ifx\paragraph\undefined\else
\let\oldparagraph\paragraph
\renewcommand{\paragraph}[1]{\oldparagraph{#1}\mbox{}}
\fi
\ifx\subparagraph\undefined\else
\let\oldsubparagraph\subparagraph
\renewcommand{\subparagraph}[1]{\oldsubparagraph{#1}\mbox{}}
\fi

\title{\texttt{exoplanet}: Gradient-based probabilistic inference for
exoplanet data \& other astronomical time series}

        \author[1]{Daniel Foreman-Mackey}
          \author[1,2]{Rodrigo Luger}
          \author[3,2]{Eric Agol}
          \author[4]{Thomas Barclay}
          \author[5]{Luke G. Bouma}
          \author[6]{Timothy D. Brandt}
          \author[7,8,9,10]{Ian Czekala}
          \author[1,11]{Trevor J. David}
          \author[7,8]{Jiayin Dong}
          \author[12]{Emily A. Gilbert}
          \author[3]{Tyler A. Gordon}
          \author[13,14]{Christina Hedges}
          \author[15,16]{Daniel R. Hey}
          \author[17]{Brett M. Morris}
          \author[1]{Adrian M. Price-Whelan}
          \author[18]{Arjun B. Savel}
    
      \affil[1]{Center for Computational Astrophysics, Flatiron
Institute, New York, NY, USA}
      \affil[2]{Virtual Planetary Laboratory, University of Washington,
Seattle, WA, USA}
      \affil[3]{Department of Astronomy, University of Washington,
University of Washington, Seattle, WA, USA}
      \affil[4]{Center for Space Sciences and Technology, University of
Maryland, Baltimore County, Baltimore, MD, USA}
      \affil[5]{Department of Astrophysical Sciences, Princeton
University, Princeton, NJ, USA}
      \affil[6]{Department of Physics, University of California, Santa
Barbara, Santa Barbara, CA, USA}
      \affil[7]{Department of Astronomy and Astrophysics, The
Pennsylvania State University, University Park, PA, USA}
      \affil[8]{Center for Exoplanets and Habitable Worlds, The
Pennsylvania State University, University Park, PA, USA}
      \affil[9]{Center for Astrostatistics, The Pennsylvania State
University, University Park, PA, USA}
      \affil[10]{Institute for Computational and Data Sciences, The
Pennsylvania State University, University Park, PA, USA}
      \affil[11]{Department of Astrophysics, American Museum of Natural
History, New York, NY, USA}
      \affil[12]{Department of Astronomy and Astrophysics, University of
Chicago, Chicago, IL, USA}
      \affil[13]{NASA Ames Research Center, Moffett Field, CA, USA}
      \affil[14]{Bay Area Environmental Research Institute, Moffett
Field, CA, USA}
      \affil[15]{Sydney Institute for Astronomy, School of Physics,
University of Sydney, Camperdown, New South Wales, Australia}
      \affil[16]{Stellar Astrophysics Centre, Department of Physics and
Astronomy, Aarhus University, Aarhus, Denmark}
      \affil[17]{Center for Space and Habitability, University of Bern,
Bern, Switzerland}
      \affil[18]{Department of Astronomy, University of Maryland,
College Park, MD, USA}
  \date{\vspace{-7ex}}

\begin{document}
\maketitle

\marginpar{

  \begin{flushleft}
  \sffamily\small

  {\bfseries DOI:} \href{https://doi.org/10.21105/joss.03285}{\color{linky}{10.21105/joss.03285}}

  \vspace{2mm}

  {\bfseries Software}
  \begin{itemize}
    \setlength\itemsep{0em}
    \item \href{https://github.com/openjournals/joss-reviews/issues/3285}{\color{linky}{Review}} \ExternalLink
    \item \href{https://github.com/exoplanet-dev/exoplanet}{\color{linky}{Repository}} \ExternalLink
    \item \href{https://doi.org/10.5281/zenodo.5006965}{\color{linky}{Archive}} \ExternalLink
  \end{itemize}

  \vspace{2mm}

  \par\noindent\hrulefill\par

  \vspace{2mm}

  {\bfseries Editor:} \href{https://arfon.org}{Arfon
Smith} \ExternalLink \\
  \vspace{1mm}
    {\bfseries Reviewers:}
  \begin{itemize}
  \setlength\itemsep{0em}
    \item \href{https://github.com/grburgess}{@grburgess}
    \item \href{https://github.com/benjaminpope}{@benjaminpope}
    \end{itemize}
    \vspace{2mm}

  {\bfseries Submitted:} 04 May 2021\\
  {\bfseries Published:} 22 June 2021

  \vspace{2mm}
  {\bfseries License}\\
  Authors of papers retain copyright and release the work under a Creative Commons Attribution 4.0 International License (\href{http://creativecommons.org/licenses/by/4.0/}{\color{linky}{CC BY 4.0}}).

  \end{flushleft}
}

\hypertarget{summary}{%
\section{Summary}\label{summary}}

\texttt{exoplanet} is a toolkit for probabilistic modeling of
astronomical time series data, with a focus on observations of
exoplanets, using \texttt{PyMC3} (Salvatier et al., 2016).
\texttt{PyMC3} is a flexible and high-performance model-building
language and inference engine that scales well to problems with a large
number of parameters. \texttt{exoplanet} extends \texttt{PyMC3}'s
modeling language to support many of the custom functions and
probability distributions required when fitting exoplanet datasets or
other astronomical time series.

While it has been used for other applications, such as the study of
stellar variability (e.g., Gillen et al., 2020; Medina et al., 2020),
the primary purpose of \texttt{exoplanet} is the characterization of
exoplanets (e.g., Gilbert et al., 2020; Plavchan et al., 2020) or
multiple-star systems (e.g., Czekala et al., 2021) using time-series
photometry, astrometry, and/or radial velocity. In particular, the
typical use case would be to use one or more of these datasets to place
constraints on the physical and orbital parameters of the system, such
as planet mass or orbital period, while simultaneously taking into
account the effects of stellar variability.

\hypertarget{statement-of-need}{%
\section{Statement of need}\label{statement-of-need}}

Time-domain astronomy is a priority of the observational astronomical
community, with huge survey datasets currently available and more
forthcoming. Within this research domain, there is significant
investment into the discovery and characterization of exoplanets,
planets orbiting stars other than our Sun. These datasets are large (on
the scale of hundreds of thousands of observations per star from
space-based observatories such as \emph{Kepler} and \emph{TESS}), and
the research questions are becoming more ambitious (in terms of both the
computational cost of the physical models and the flexibility of these
models). The packages in the \emph{exoplanet} ecosystem are designed to
enable rigorous probabilistic inference with these large datasets and
high-dimensional models by providing a high-performance and well-tested
infrastructure for integrating these models with modern modeling
frameworks such as \texttt{PyMC3}. Since its initial release at the end
of 2018, \texttt{exoplanet} has been widely used, with 64 citations of
the Zenodo record (Foreman-Mackey et al., 2021) so far.

\hypertarget{the-exoplanet-software-ecosystem}{%
\section{\texorpdfstring{The \emph{exoplanet} software
ecosystem}{The exoplanet software ecosystem}}\label{the-exoplanet-software-ecosystem}}

Besides the primary \texttt{exoplanet} package, the \emph{exoplanet}
ecosystem of projects includes several other libraries. This paper
describes, and is the primary reference for, this full suite of
packages. The following provides a short description of each library
within this ecosystem and discusses how they are related.

\begin{itemize}
\tightlist
\item
  \texttt{exoplanet}\footnote{\url{https://github.com/exoplanet-dev/exoplanet}}
  is the primary library, and it includes implementations of many
  special functions required for exoplanet data analysis. These include
  the spherical geometry for computing orbits, some exoplanet-specific
  distributions for eccentricity (Kipping, 2013a; Van Eylen et al.,
  2019) and limb darkening (Kipping, 2013b), and exposure-time
  integrated limb-darkened transit light curves.
\item
  \texttt{exoplanet-core}\footnote{\url{https://github.com/exoplanet-dev/exoplanet-core}}
  provides efficient, well-tested, and differentiable implementations of
  all of the exoplanet-specific operations that must be compiled for
  performance. These include an efficient solver for Kepler's equation
  (based on the algorithm proposed by Raposo-Pulido \& Peláez, 2017) and
  limb darkened transit light curves (Agol et al., 2020). Besides the
  implementation for \texttt{PyMC3}, \texttt{exoplanet-core} includes
  implementations in \texttt{numpy} (Harris et al., 2020) and
  \texttt{jax} (Bradbury et al., 2018).
\item
  \texttt{celerite2}\footnote{\url{https://celerite2.readthedocs.io}},
  is an updated implementation of the \emph{celerite}
  algorithm\footnote{\url{https://celerite.readthedocs.io}}
  (Foreman-Mackey, 2018; Foreman-Mackey et al., 2017) for scalable
  Gaussian Process regression for time series data. Like
  \texttt{exoplanet-core}, \texttt{celerite2} includes support for
  \texttt{numpy}, \texttt{jax}, and \texttt{PyMC3}, as well as some
  recent generalizations of the \emph{celerite} algorithm (Gordon et
  al., 2020).
\item
  \texttt{pymc3-ext}\footnote{\url{https://github.com/exoplanet-dev/pymc3-ext}},
  includes a set of helper functions to make \texttt{PyMC3} more
  amenable to the typical astronomical data analysis workflow. For
  example, it provides a tuning schedule for \texttt{PyMC3}'s sampler
  (based on the method used by the \texttt{Stan} project and described
  by Carpenter et al., 2017) that provides better performance on models
  with correlated parameters.
\item
  \texttt{rebound-pymc3}\footnote{\url{https://github.com/exoplanet-dev/rebound-pymc3}}
  provides an interface between \emph{REBOUND} (Rein \& Liu, 2012),
  \emph{REBOUNDx} (Tamayo et al., 2020), and \texttt{PyMC3} to enable
  inference with full N-body orbit integration.
\end{itemize}

\hypertarget{documentation-case-studies}{%
\section{Documentation \& case
studies}\label{documentation-case-studies}}

The main documentation page for the \emph{exoplanet} libraries lives at
\href{https://docs.exoplanet.codes}{docs.exoplanet.codes} where it is
hosted on \href{https://readthedocs.org}{ReadTheDocs}. The tutorials
included with the documentation are automatically executed on every push
or pull request to the GitHub repository, with the goal of ensuring that
the tutorials are always compatible with the current version of the
code. The \texttt{celerite2} project has its own documentation page at
\href{https://celerite2.readthedocs.io}{celerite2.readthedocs.io}, with
tutorials that are similarly automatically executed.

\begin{figure}
\centering
\includegraphics{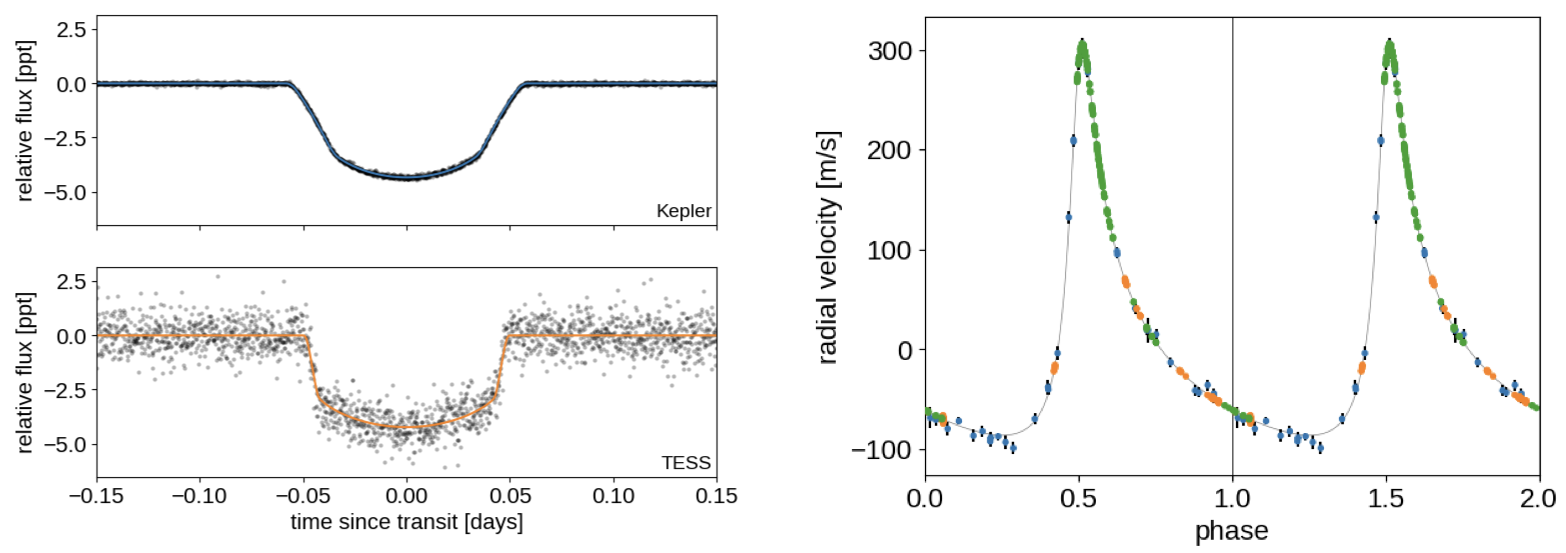}
\caption{Some examples of datasets fit using \texttt{exoplanet}. The
full analyses behind these examples are available on the ``Case
Studies'' page as Jupyter notebooks. (left) A fit to the light curves of
a transiting exoplanet observed by two different space-based photometric
surveys: Kepler and TESS. (right) The phase-folded radial velocity time
series for an exoplanet observed from different observatories with
different instruments, fit simultaneously using \texttt{exoplanet}.
\label{fig:figure}}
\end{figure}

Alongside these documentation pages, there is a parallel ``Case
Studies'' website at
\href{https://gallery.exoplanet.codes}{gallery.exoplanet.codes} that
includes more detailed example use cases for \texttt{exoplanet} and the
other libraries described here. Like the tutorials on the documentation
page, these case studies are automatically executed using GitHub
Actions, but at lower cadence (once a week and when a new release of the
\texttt{exoplanet} library is made) since the runtime is much longer.
\autoref{fig:figure} shows the results of two example case studies
demonstrating some of the potential use cases of the \texttt{exoplanet}
software ecosystem.

\hypertarget{similar-tools}{%
\section{Similar tools}\label{similar-tools}}

There is a rich ecosystem of tooling available for inference with models
such as the ones supported by \texttt{exoplanet}. Each of these tools
has its own set of strengths and limitations and we will not make a
detailed comparison here, but it is worth listing some of these tools
and situating \texttt{exoplanet} in this context.

Some of the most popular tools in this space include (and note that this
is far from a comprehensive list!) \texttt{EXOFAST} (J. Eastman et al.,
2013; J. D. Eastman et al., 2019), \texttt{radvel} (Fulton et al.,
2018), \texttt{juliet} (Espinoza et al., 2019), \texttt{exostriker}
(Trifonov, 2019), \texttt{PYANETI} (Barragán et al., 2019),
\texttt{allesfitter} (Günther \& Daylan, 2021), and \texttt{orbitize}
(Blunt et al., 2020). Similar tools also exist for modeling observations
of eclipsing binary systems, including \texttt{JKTEBOP} (Southworth et
al., 2004), \texttt{eb} (Irwin et al., 2011), and \texttt{PHOEBE}
(Conroy et al., 2020). These packages all focus on providing a
high-level interface for designing models and then executing a fit. In
contrast, \texttt{exoplanet} is designed to be lower level and more
conceptually similar to tools like \texttt{batman} (Kreidberg, 2015),
\texttt{PyTransit} (Hannu Parviainen, 2015), \texttt{ldtk} (H.
Parviainen \& Aigrain, 2015), \texttt{ellc} (Maxted, 2016),
\texttt{starry} (Luger et al., 2019), or \texttt{Limbdark.jl} (Agol et
al., 2020), which provide the building blocks for evaluating the models
required for inference with exoplanet datasets. In fact, several of the
higher-level packages listed above include these lower-level libraries
as dependencies, and our hope is that \texttt{exoplanet} could provide
the backend for future high-level libraries.

As emphasized in the title of this paper, the main selling point of
\texttt{exoplanet} when compared to other tools in this space is that it
supports differentiation of all components of the model and is designed
to integrate seamlessly with the \texttt{aesara} (Willard et al., 2021)
automatic differentiation framework used by \texttt{PyMC3}. It is worth
noting that \texttt{aesara} was previously known as \texttt{Theano}
(Theano Development Team, 2016), so these names are sometimes used
interchangeably in the \texttt{PyMC3} or \texttt{exoplanet}
documentation\footnote{More information about this distinction is
  available at \url{https://docs.exoplanet.codes/en/stable/user/theano/}}.
This allows the use of modern inference algorithms such as No U-Turn
Sampling (Hoffman \& Gelman, 2014) or Automatic Differentiation
Variational Inference (Kucukelbir et al., 2017). These algorithms can
have some computational and conceptual advantages over inference methods
that do not use gradients, especially for high-dimensional models. The
computation of gradients is also useful for model optimization; this is
necessary when, say, searching for new exoplanets, mapping out
degeneracies or multiple modes of a posterior, or estimating
uncertainties from a Hessian. Care has been taken to provide gradients
which are numerically stable, and more accurate and faster to evaluate
than finite-difference gradients.

\hypertarget{acknowledgements}{%
\section{Acknowledgements}\label{acknowledgements}}

We would like to thank the Astronomical Data Group at Flatiron for
listening to every iteration of this project and for providing great
feedback every step of the way.

This research was partially conducted during the \emph{Exostar19}
program at the \emph{Kavli Institute for Theoretical Physics} at UC
Santa Barbara, which was supported in part by the National Science
Foundation under Grant No.~NSF PHY-1748958.

Besides the software cited above, \texttt{exoplanet} is also built on
top of \texttt{ArviZ} (Kumar et al., 2019) and \texttt{AstroPy} (Astropy
Collaboration et al., 2018, 2013).

\hypertarget{references}{%
\section*{References}\label{references}}
\addcontentsline{toc}{section}{References}

\hypertarget{refs}{}
\begin{CSLReferences}{1}{0}
\leavevmode\hypertarget{ref-agol20}{}%
Agol, E., Luger, R., \& Foreman-Mackey, D. (2020). {Analytic Planetary
Transit Light Curves and Derivatives for Stars with Polynomial Limb
Darkening}. \emph{The Astronomical Journal}, \emph{159}(3), 123.
\url{https://doi.org/10.3847/1538-3881/ab4fee}

\leavevmode\hypertarget{ref-astropy18}{}%
Astropy Collaboration, Price-Whelan, A. M., Sipőcz, B. M., Günther, H.
M., Lim, P. L., Crawford, S. M., Conseil, S., Shupe, D. L., Craig, M.
W., Dencheva, N., Ginsburg, A., VanderPlas, J. T., Bradley, L. D.,
Pérez-Suárez, D., de Val-Borro, M., Aldcroft, T. L., Cruz, K. L.,
Robitaille, T. P., Tollerud, E. J., \ldots{} Astropy Contributors.
(2018). {The Astropy Project: Building an Open-science Project and
Status of the v2.0 Core Package}. \emph{The Astronomical Journal},
\emph{156}, 123. \url{https://doi.org/10.3847/1538-3881/aabc4f}

\leavevmode\hypertarget{ref-astropy13}{}%
Astropy Collaboration, Robitaille, T. P., Tollerud, E. J., Greenfield,
P., Droettboom, M., Bray, E., Aldcroft, T., Davis, M., Ginsburg, A.,
Price-Whelan, A. M., Kerzendorf, W. E., Conley, A., Crighton, N.,
Barbary, K., Muna, D., Ferguson, H., Grollier, F., Parikh, M. M., Nair,
P. H., \ldots{} Streicher, O. (2013). {Astropy: A community Python
package for astronomy}. \emph{Astronomy \& Astrophysics}, \emph{558},
A33. \url{https://doi.org/10.1051/0004-6361/201322068}

\leavevmode\hypertarget{ref-barragan19}{}%
Barragán, O., Gandolfi, D., \& Antoniciello, G. (2019). {PYANETI: a fast
and powerful software suite for multiplanet radial velocity and transit
fitting}. \emph{Monthly Notices of the Royal Astronomical Society},
\emph{482}, 1017--1030. \url{https://doi.org/10.1093/mnras/sty2472}

\leavevmode\hypertarget{ref-blunt20}{}%
Blunt, S., Wang, J. J., Angelo, I., Ngo, H., Cody, D., De Rosa, R. J.,
Graham, J. R., Hirsch, L., Nagpal, V., Nielsen, E. L., Pearce, L., Rice,
M., \& Tejada, R. (2020). {orbitize!: A Comprehensive Orbit-fitting
Software Package for the High-contrast Imaging Community}. \emph{The
Astronomical Journal}, \emph{159}(3), 89.
\url{https://doi.org/10.3847/1538-3881/ab6663}

\leavevmode\hypertarget{ref-jax}{}%
Bradbury, J., Frostig, R., Hawkins, P., Johnson, M. J., Leary, C.,
Maclaurin, D., Necula, G., Paszke, A., VanderPlas, J., Wanderman-Milne,
S., \& Zhang, Q. (2018). \emph{{JAX}: Composable transformations of
{P}ython+{N}um{P}y programs} (Version 0.2.5) {[}Computer software{]}.
\url{http://github.com/google/jax}

\leavevmode\hypertarget{ref-carpenter17}{}%
Carpenter, B., Gelman, A., Hoffman, M. D., Lee, D., Goodrich, B.,
Betancourt, M., Brubaker, M., Guo, J., Li, P., \& Riddell, A. (2017).
{Stan: A Probabilistic Programming Language}. \emph{Journal of
Statistical Software}, \emph{76}(1), 1--32.
\url{https://doi.org/10.18637/jss.v076.i01}

\leavevmode\hypertarget{ref-conroy20}{}%
Conroy, K. E., Kochoska, A., Hey, D., Pablo, H., Hambleton, K. M.,
Jones, D., Giammarco, J., Abdul-Masih, M., \& Prša, A. (2020). {Physics
of Eclipsing Binaries. V. General Framework for Solving the Inverse
Problem}. \emph{The Astrophysical Journal Supplement Series},
\emph{250}(2), 34. \url{https://doi.org/10.3847/1538-4365/abb4e2}

\leavevmode\hypertarget{ref-czekala21}{}%
Czekala, I., Ribas, Á., Cuello, N., Chiang, E., Macías, E., Duchêne, G.,
Andrews, S. M., \& Espaillat, C. C. (2021). {A Coplanar Circumbinary
Protoplanetary Disk in the TWA 3 Triple M Dwarf System}. \emph{The
Astrophysical Journal}, \emph{912}(1), 6.
\url{https://doi.org/10.3847/1538-4357/abebe3}

\leavevmode\hypertarget{ref-eastman19}{}%
Eastman, J. D., Rodriguez, J. E., Agol, E., Stassun, K. G., Beatty, T.
G., Vanderburg, A., Gaudi, B. S., Collins, K. A., \& Luger, R. (2019).
{EXOFASTv2: A public, generalized, publication-quality exoplanet
modeling code}. \emph{arXiv e-Prints}, arXiv:1907.09480.
\url{http://arxiv.org/abs/1907.09480}

\leavevmode\hypertarget{ref-eastman13}{}%
Eastman, J., Gaudi, B. S., \& Agol, E. (2013). {EXOFAST: A Fast
Exoplanetary Fitting Suite in IDL}. \emph{Publications of the
Astronomical Society of the Pacific}, \emph{125}(923), 83.
\url{https://doi.org/10.1086/669497}

\leavevmode\hypertarget{ref-espinoza19}{}%
Espinoza, N., Kossakowski, D., \& Brahm, R. (2019). {juliet: a versatile
modelling tool for transiting and non-transiting exoplanetary systems}.
\emph{Monthly Notices of the Royal Astronomical Society}, \emph{490}(2),
2262--2283. \url{https://doi.org/10.1093/mnras/stz2688}

\leavevmode\hypertarget{ref-foremanmackey18}{}%
Foreman-Mackey, D. (2018). {Scalable Backpropagation for Gaussian
Processes using Celerite}. \emph{Research Notes of the American
Astronomical Society}, \emph{2}(1), 31.
\url{https://doi.org/10.3847/2515-5172/aaaf6c}

\leavevmode\hypertarget{ref-foremanmackey17}{}%
Foreman-Mackey, D., Agol, E., Ambikasaran, S., \& Angus, R. (2017).
{Fast and Scalable Gaussian Process Modeling with Applications to
Astronomical Time Series}. \emph{The Astronomical Journal}, \emph{154},
220. \url{https://doi.org/10.3847/1538-3881/aa9332}

\leavevmode\hypertarget{ref-zenodo}{}%
Foreman-Mackey, D., Savel, A., Luger, R., Czekala, I., Agol, E.,
Price-Whelan, A., Hedges, C., Gilbert, E., Barclay, T., Bouma, L., \&
Brandt, T. D. (2021). \emph{{exoplanet-dev/exoplanet}} (Version 0.4.5)
{[}Computer software{]}. Zenodo.
\url{https://doi.org/10.5281/zenodo.1998447}

\leavevmode\hypertarget{ref-fulton18}{}%
Fulton, B. J., Petigura, E. A., Blunt, S., \& Sinukoff, E. (2018).
{RadVel: The Radial Velocity Modeling Toolkit}. \emph{Publications of
the Astronomical Society of the Pacific}, \emph{130}(986), 044504.
\url{https://doi.org/10.1088/1538-3873/aaaaa8}

\leavevmode\hypertarget{ref-gilbert20}{}%
Gilbert, E. A., Barclay, T., Schlieder, J. E., Quintana, E. V., Hord, B.
J., Kostov, V. B., Lopez, E. D., Rowe, J. F., Hoffman, K., Walkowicz, L.
M., Silverstein, M. L., Rodriguez, J. E., Vanderburg, A., Suissa, G.,
Airapetian, V. S., Clement, M. S., Raymond, S. N., Mann, A. W., Kruse,
E., \ldots{} Winters, J. G. (2020). {The First Habitable-zone
Earth-sized Planet from TESS. I. Validation of the TOI-700 System}.
\emph{The Astronomical Journal}, \emph{160}(3), 116.
\url{https://doi.org/10.3847/1538-3881/aba4b2}

\leavevmode\hypertarget{ref-gillen20}{}%
Gillen, E., Briegal, J. T., Hodgkin, S. T., Foreman-Mackey, D., Van
Leeuwen, F., Jackman, J. A. G., McCormac, J., West, R. G., Queloz, D.,
Bayliss, D., Goad, M. R., Watson, C. A., Wheatley, P. J., Belardi, C.,
Burleigh, M. R., Casewell, S. L., Jenkins, J. S., Raynard, L., Smith, A.
M. S., \ldots{} Vines, J. I. (2020). {NGTS clusters survey - I. Rotation
in the young benchmark open cluster Blanco 1}. \emph{Monthly Notices of
the Royal Astronomical Society}, \emph{492}(1), 1008--1024.
\url{https://doi.org/10.1093/mnras/stz3251}

\leavevmode\hypertarget{ref-gordon20}{}%
Gordon, T. A., Agol, E., \& Foreman-Mackey, D. (2020). {A Fast,
Two-dimensional Gaussian Process Method Based on Celerite: Applications
to Transiting Exoplanet Discovery and Characterization}. \emph{The
Astronomical Journal}, \emph{160}(5), 240.
\url{https://doi.org/10.3847/1538-3881/abbc16}

\leavevmode\hypertarget{ref-guenther20}{}%
Günther, M. N., \& Daylan, T. (2021). {Allesfitter: Flexible Star and
Exoplanet Inference from Photometry and Radial Velocity}. \emph{The
Astrophysical Journal Supplement Series}, \emph{254}(1), 13.
\url{https://doi.org/10.3847/1538-4365/abe70e}

\leavevmode\hypertarget{ref-numpy}{}%
Harris, C. R., Millman, K. J., Walt, S. J. van der, Gommers, R.,
Virtanen, P., Cournapeau, D., Wieser, E., Taylor, J., Berg, S., Smith,
N. J., Kern, R., Picus, M., Hoyer, S., Kerkwijk, M. H. van, Brett, M.,
Haldane, A., Río, J. F. del, Wiebe, M., Peterson, P., \ldots{} Oliphant,
T. E. (2020). Array programming with {NumPy}. \emph{Nature},
\emph{585}(7825), 357--362.
\url{https://doi.org/10.1038/s41586-020-2649-2}

\leavevmode\hypertarget{ref-hoffman14}{}%
Hoffman, M. D., \& Gelman, A. (2014). {The No-U-Turn sampler: adaptively
setting path lengths in Hamiltonian Monte Carlo}. \emph{Journal of
Machine Learning Research}, \emph{15}(1), 1593--1623.
\url{https://jmlr.org/papers/v15/hoffman14a.html}

\leavevmode\hypertarget{ref-irwin11}{}%
Irwin, J. M., Quinn, S. N., Berta, Z. K., Latham, D. W., Torres, G.,
Burke, C. J., Charbonneau, D., Dittmann, J., Esquerdo, G. A., Stefanik,
R. P., Oksanen, A., Buchhave, L. A., Nutzman, P., Berlind, P., Calkins,
M. L., \& Falco, E. E. (2011). {LSPM J1112+7626: Detection of a 41 Day
M-dwarf Eclipsing Binary from the MEarth Transit Survey}. \emph{The
Astrophysical Journal}, \emph{742}(2), 123.
\url{https://doi.org/10.1088/0004-637X/742/2/123}

\leavevmode\hypertarget{ref-kipping13b}{}%
Kipping, D. M. (2013a). {Parametrizing the exoplanet eccentricity
distribution with the beta distribution.} \emph{Monthly Notices of the
Royal Astronomical Society}, \emph{434}, L51--L55.
\url{https://doi.org/10.1093/mnrasl/slt075}

\leavevmode\hypertarget{ref-kipping13}{}%
Kipping, D. M. (2013b). {Efficient, uninformative sampling of limb
darkening coefficients for two-parameter laws}. \emph{Monthly Notices of
the Royal Astronomical Society}, \emph{435}, 2152--2160.
\url{https://doi.org/10.1093/mnras/stt1435}

\leavevmode\hypertarget{ref-kreidberg15}{}%
Kreidberg, L. (2015). {batman: BAsic Transit Model cAlculatioN in
Python}. \emph{Publications of the Astronomical Society of the Pacific},
\emph{127}(957), 1161. \url{https://doi.org/10.1086/683602}

\leavevmode\hypertarget{ref-kucukelbir17}{}%
Kucukelbir, A., Tran, D., Ranganath, R., Gelman, A., \& Blei, D. M.
(2017). {Automatic Differentiation Variational Inference}. \emph{Journal
of Machine Learning Research}, \emph{18}(14), 1--45.
\url{http://jmlr.org/papers/v18/16-107.html}

\leavevmode\hypertarget{ref-arviz}{}%
Kumar, R., Carroll, C., Hartikainen, A., \& Martin, O. A. (2019).
{ArviZ} a unified library for exploratory analysis of {Bayesian} models
in {Python}. \emph{The Journal of Open Source Software}.
\url{https://doi.org/10.21105/joss.01143}

\leavevmode\hypertarget{ref-luger19}{}%
Luger, R., Agol, E., Foreman-Mackey, D., Fleming, D. P., Lustig-Yaeger,
J., \& Deitrick, R. (2019). {starry: Analytic Occultation Light Curves}.
\emph{The Astronomical Journal}, \emph{157}, 64.
\url{https://doi.org/10.3847/1538-3881/aae8e5}

\leavevmode\hypertarget{ref-maxted16}{}%
Maxted, P. F. L. (2016). {ellc: A fast, flexible light curve model for
detached eclipsing binary stars and transiting exoplanets}.
\emph{Astronomy \& Astrophysics}, \emph{591}, A111.
\url{https://doi.org/10.1051/0004-6361/201628579}

\leavevmode\hypertarget{ref-medina20}{}%
Medina, A. A., Winters, J. G., Irwin, J. M., \& Charbonneau, D. (2020).
{Flare Rates, Rotation Periods, and Spectroscopic Activity Indicators of
a Volume-complete Sample of Mid- to Late-M Dwarfs within 15 pc}.
\emph{The Astrophysical Journal}, \emph{905}(2), 107.
\url{https://doi.org/10.3847/1538-4357/abc686}

\leavevmode\hypertarget{ref-parviainen15}{}%
Parviainen, Hannu. (2015). {PYTRANSIT: fast and easy exoplanet transit
modelling in PYTHON}. \emph{Monthly Notices of the Royal Astronomical
Society}, \emph{450}(3), 3233--3238.
\url{https://doi.org/10.1093/mnras/stv894}

\leavevmode\hypertarget{ref-parviainen15b}{}%
Parviainen, H., \& Aigrain, S. (2015). {LDTK: Limb Darkening Toolkit}.
\emph{Monthly Notices of the Royal Astronomical Society}, \emph{453}(4),
3821--3826. \url{https://doi.org/10.1093/mnras/stv1857}

\leavevmode\hypertarget{ref-plavchan20}{}%
Plavchan, P., Barclay, T., Gagné, J., Gao, P., Cale, B., Matzko, W.,
Dragomir, D., Quinn, S., Feliz, D., Stassun, K., Crossfield, I. J. M.,
Berardo, D. A., Latham, D. W., Tieu, B., Anglada-Escudé, G., Ricker, G.,
Vanderspek, R., Seager, S., Winn, J. N., \ldots{} Zilberman, P. (2020).
{A planet within the debris disk around the pre-main-sequence star AU
Microscopii}. \emph{Nature}, \emph{582}(7813), 497--500.
\url{https://doi.org/10.1038/s41586-020-2400-z}

\leavevmode\hypertarget{ref-raposo17}{}%
Raposo-Pulido, V., \& Peláez, J. (2017). {An efficient code to solve the
Kepler equation. Elliptic case}. \emph{Monthly Notices of the Royal
Astronomical Society}, \emph{467}(2), 1702--1713.
\url{https://doi.org/10.1093/mnras/stx138}

\leavevmode\hypertarget{ref-rein12}{}%
Rein, H., \& Liu, S.-F. (2012). {REBOUND: an open-source multi-purpose
N-body code for collisional dynamics}. \emph{Astronomy \& Astrophysics},
\emph{537}, A128. \url{https://doi.org/10.1051/0004-6361/201118085}

\leavevmode\hypertarget{ref-pymc3}{}%
Salvatier, J., Wiecki, T. V., \& Fonnesbeck, C. (2016). Probabilistic
programming in python using PyMC3. \emph{PeerJ Computer Science},
\emph{2}, e55. \url{https://doi.org/10.7717/peerj-cs.55}

\leavevmode\hypertarget{ref-southworth04}{}%
Southworth, J., Maxted, P. F. L., \& Smalley, B. (2004). {Eclipsing
binaries in open clusters - II. V453 Cyg in NGC 6871}. \emph{Monthly
Notices of the Royal Astronomical Society}, \emph{351}(4), 1277--1289.
\url{https://doi.org/10.1111/j.1365-2966.2004.07871.x}

\leavevmode\hypertarget{ref-tamayo20}{}%
Tamayo, D., Rein, H., Shi, P., \& Hernandez, D. M. (2020). {REBOUNDx: a
library for adding conservative and dissipative forces to otherwise
symplectic N-body integrations}. \emph{Monthly Notices of the Royal
Astronomical Society}, \emph{491}(2), 2885--2901.
\url{https://doi.org/10.1093/mnras/stz2870}

\leavevmode\hypertarget{ref-theano}{}%
Theano Development Team. (2016). {Theano: A {Python} framework for fast
computation of mathematical expressions}. \emph{arXiv e-Prints},
\emph{abs/1605.02688}. \url{http://arxiv.org/abs/1605.02688}

\leavevmode\hypertarget{ref-trifonov19}{}%
Trifonov, T. (2019). \emph{{The Exo-Striker: Transit and radial velocity
interactive fitting tool for orbital analysis and N-body simulations}}
(p. ascl:1906.004).

\leavevmode\hypertarget{ref-vaneylen19}{}%
Van Eylen, V., Albrecht, S., Huang, X., MacDonald, M. G., Dawson, R. I.,
Cai, M. X., Foreman-Mackey, D., Lundkvist, M. S., Silva Aguirre, V.,
Snellen, I., \& Winn, J. N. (2019). {The Orbital Eccentricity of Small
Planet Systems}. \emph{The Astronomical Journal}, \emph{157}(2), 61.
\url{https://doi.org/10.3847/1538-3881/aaf22f}

\leavevmode\hypertarget{ref-aesara}{}%
Willard, B. T., Osthege, M., Ho, G., Vieira, R., Wiecki, T.,
Foreman-Mackey, D., Chaudhari, K., Legrand, N., Kumar, R., Lao, J.,
Abril-Pla, O., Fonnesbeck, C., Goldman, R. P., \& Gorelli, M. (2021).
\emph{{pymc-devs/aesara}} (Version 2.0.7) {[}Computer software{]}.
Zenodo. \url{https://doi.org/10.5281/zenodo.4695331}

\end{CSLReferences}

\end{document}